Theoretical prediction of sandwiched two-dimensional phosphide binary compounds sheets with tunable bandgaps and anisotropic physical properties

*C. Y. Zhang and M. Yu\**

Department of Physics and Astronomy, University of Louisville, Louisville, KY 40292

\* Correspondence author: Ming Yu (m0yu0001@louisville.edu)

**Abstract**

Atomic layers of GaP and InP binary compounds with unique anisotropic structural, electronic, and mechanical properties have been predicted from the first principle molecular dynamics simulations. These new members of phosphide binary compounds family stabilize to a sandwiched two-dimensional (2D) crystalline structure with orthorhombic lattices symmetry and high buckling of 2.14 Å-2.46 Å. Their vibration modes are similar to those of phosphorene with six Raman active modes ranging from ~80 $cm^{-1}$ to 400 $cm^{-1}$. The speeds of sound in their phono dispersions reflect anisotropy in their elastic constants, which was further confirmed from their strong directional dependence of Young's moduli and effective nonlinear elastic moduli. They show wide bandgap semiconductor behavior with fundamental bandgaps of 2.89 eV for GaP and 2.59 eV for InP, respectively, even wider than their bulk counterparts. Such bandgaps were found to be tunable under the strains. In particular, a direct-indirect bandgap transition was found under certain strains along zigzag or biaxial orientations, reflecting their promising applications for the strain-induced bandgap engineering in nanoelectronics and photovoltaics. Feasible pathways to realize these novel 2D phosphide compounds were also proposed in this work.

PACS: 61.46.-w, 68.65.-k, 73.22.-f, 73.61.Ey, 81.05.Zx

## 1. Introduction

Exploring unknown two-dimensional (2D) nanomaterials that have targeted physical properties for nanoscale electronic and optoelectronic devices is urgently demanded in 2D nanomaterial research in the post-graphene era [1]. Tremendous interests were mainly focused on the discovery of analogous 2D nanomaterials from their layered bulk counterparts (*i.e.*, those with Van der Waals interaction between layers). Mechanical exfoliation and chemical vapor deposition (CVD) have been successfully used in realizing these types of 2D materials [2], including the hexagonal boron-nitride (*h*-BN) sheet [3, 4], transition metal dichalcogenides (TMDs) layered structures with $MX_2$ grouping (*e.g.*, M=Mo, W and X = S, Se, Te) [5-7], and phosphorene [8-10]. Those newly discovered 2D materials were energetically preferential and indeed show their unique electronic, optical, chemical, and mechanical properties, indicating their remarkable promising applications for nanoelectronics, optics, catalysts, etc. [6-10]. Extensive interests were also focused on the discovery of other possible types of 2D nanomaterials which have no corresponding layered bulk counterpatrts in nature, and mechanical stripping or general CVD method might not work easily to realize those types of materials. It is extremly tough and challenge to discover these types of 2D nanomaterials. Recently, large amount of efforts have been devoted on this issue both theoretically and experimentally [11-22] One of the big breakthroughs was the prediction of silicene and germanene [16-16, 23-34]. Theoretical calculations pointed out that they are dynamically stable as free standing sheets with low buckled honeycomb lattice structures (*i.e.*, the two sublattices are relatively shifted in the direction perpendicular to the atomic plane with buckling of 0.44 Å for silicene and 0.67 Å for germanene, respectively) [17]. Several experimental results found that, instead of free standing sheets, they can be realized on metal substracts [25-34] Borophene, another new member of elemental monolayers with various patterns in structure, was also theoretically predicted quite recently [35, 36], and has been successfully synthesized on the Ag (111) surface with novel properties of Dirac fermions [37-40].

These new discoveries open a door to further explore unknown 2D materials, in particular, the unkonwn 2D binary compounds such as SiC, GeC, and III-V binary compounds [15, 19-22]. Low buckled honeycomb InP and GaP monolayers were first predicted by H. Sahin *et al*. [19] when they studied monolayer honeycomb structures of group-IV elements and III-V binary compounds. Their density functional theory (DFT) calculations showed that those low buckled honeycomb InP and GaP binary compounds have indirect energy bandgaps in the range of 1~2 eV, indicating their possible applications for optoelectronic devices. However, it is not clear, whether these low buckled honeycomb structures are the only form of the 2D GaP/InP binary compound sheets, and if there exist any other allotropes of 2D GaP/InP binary compounds sheets which are even energetically more stable.

In this paper, we report our recent systematic study on seeking stable and energetically preferential 2D monalayers of GaP and InP binary compounds sheets. Our first principle molecular dynamics simulations show that, in addition to the low buckled honeycomb structures, a new 2D monolayer structure can be obtained by bulk truncation along a special orientation, called 'armchair truncation'. These newly discovered allotropes of 2D GaP/InP binary compounds possess high puckered and sandwiched monolayer structures with orthorhombic lattice symmetry, and are energetically much stable than the previously predicted[19] low buckled honeycomb GaP/InP sheets. More interesting, they possess strong anisotropic electronic and mechanical properties. Their fundamental bandgaps are wider than those of low buckled honeycomb sheets and even wider than their bulk counterparts. Such bandgaps are found to be tunable under the strain along armchair/zigzag direction, and a transition from the indirect to the direct band gap could occur along particular orientations.

## 2. Computational details

In the processes of seeking new allotropes of 2D monolayer GaP/InP binary compounds, we employed the DFT [41, 42] framework, as implemented in the Vienna Ab-initio Simulation Package (VASP) [43], and performed the structure optimization, dynamic stability analysis, and electronic and mechanical properties calculations. The electron-ion interactions were described by the Projector Augmented Wave (PAW) [44], while electron exchange-correlation interactions were treated by the generalized gradient approximation (GGA) [45] in the scheme of Perdew Burke Ernzerhof (PBE) [46]. The structural relaxation was performed using Congregate-Gradient algorithm [47] implemented in VASP. The periodic boundary conditions were chosen in the layered plane with a vacuum space of 15 Å between adjacent layers to avoid any mirror interactions. An energy cutoff was set to be 500 eV for the plane wave basis in all calculations, and the criteria for the convergences of energy and force in relaxation processes were set to be $10^{-5}$ eV and $10^{-4}$ eV/Å, respectively. A 1x1 rectangular primitive cell was chosen to study the 2D GaP and InP monolayer structures, and the Brillouin zones (BZ) were sampled by $25 \times 25 \times 1$ k-point meshes generated in accordance with the Monkhorst-Pack scheme [48] in the optimization and band structure calculations.

A benchmark for zinc blende bulk GaP/InP crystalline structures was carried out (see the last columns of Tables 1-1 and 1-2). The optimized lattice constants are 5.53 Å for bulk GaP and 6.02 Å for bulk InP, respectively, which are only ~ 1.47% for GaP and 2.38% for InP overestimated compared to the experimental measurements [49]. Calculated cohesive energies (*i.e.*, -8.54 eV/pair for GaP and -7.78 eV/pair for InP, respectively), on the other hand, are about 1 eV per pair lower than the experimental values [50], which are typical in DFT-GGA calculations. It is also common that calculated DFT band gaps (*i.e.*, ~1.52 eV for GaP and ~0.38 eV for InP, respectively) are ~ 1 eV underestimated compared with the experimental results (*i.e.*, 2.26 eV for GaP and 1.34 eV for InP [49]), mostly due to the lack of self-energies corrections in DFT calculations. Such big errors in the DFT band gap calculations were reduced using Heyd–Scuseria–Ernzerhof screened Coulomb hybrid functional [51], as implemented in VASP (referred as HSE06). The calculated HSE06 band gaps are 2.37 eV for bulk GaP and 1.10 eV for bulk InP, resulting in a mean absolute error of only ~ 0.11 eV and ~0.24 eV, as compared with the experimental measurements.

### 3. Results and Discussions

### 3.1 Anisotropic crystalline structures

The most stable phase of GaP/InP binary compounds in the nature is the zinc blende crystalline structure, followed by the wurtzite structure. In both phases, Ga (In) and P atoms prefer to form $sp^3$ type of hybrid orbitals. Based on this chemical bonding nature, we proposed to search the possible existence of unknown allotropes of 2D GaP/InP binary compounds by truncating the bulk GaP/InP, *e.g.,* the zinc blende structure, along certain orientations. As illustrated in Fig. 1 (a), the zinc blende crystalline structures of GaP and InP binary compounds along (111) orientation can be viewed as a series of bilayers aligned with ABC stacking sequences and interacted by strong Coulomb interactions. When a zigzag truncation is performed (shown by the red-dashed box in Fig. 1 (a)), a buckled bilayer sheet was constructed (shown in Fig. 1 (b)). During the structural relaxation, this initial configuration was then stabilized to a so called low buckled honeycomb structure with the buckling of 0.36 and 0.54 Å for GaP and InP sheets, respectively, consistent with the previous results predicted by H. Şahin *et al*. [19]. Alternatively, when an armchair truncation is performed (shown by the blue-dashed box in Fig. 1 (a)), a puckered bilayer sheet, analogous to phosphorene, was constructed (shown in Fig. 1 (c)). Very interesting, such initial configuration underwent a structure transition during the structural relaxation.

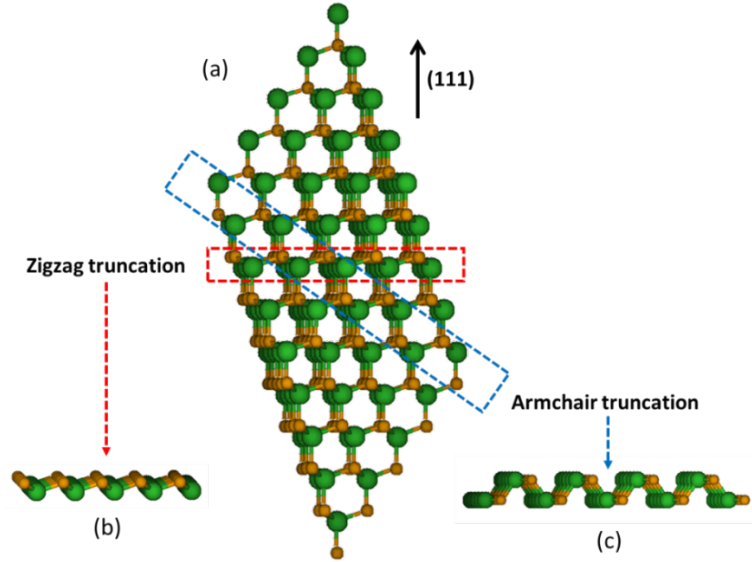

**Figure 1.** (color online) Schematic illustrations of the zinc blende GaP/InP crystalline structures along (111) orientation (a), a bilayer sheet truncated along the zigzag direction (b) (indicated by the red-dashed box in (a)), and a bilayer sheet truncated along the armchair direction (c) (indicated by the blue-dashed box in (a)). The yellow balls represent P atoms, and the green balls, the Ga/In atoms, respectively.

The corresponding molecular dynamics process is illustrated in Figs. 2 (a)-(c). It is found that Ga/In atoms initially located on the ridge and in the valley (shown in Fig. 2 (a)) move towards each other vertically during the relaxation (shown in Fig. 2 (b)), and finally, the bilayer sheet stabilized to a highly puckered structure, where Ga/In atoms are sandwiched by P atoms and form a semi-chair shape of hexagons together with P atoms. The top and side views of such stabilized structure are shown in Fig. 2 (c). Different from the low buckled honeycomb 2D GaP/InP with rhombohedral lattice symmetry and three-fold rotation symmetry $C_3$, the new 2D allotropes of GaP/InP binary compound possess orthorhombic lattice symmetry which belong to the space group of P11m and point group of $C_s^1$, reflecting the anisotropic feature along the zigzag and armchair directions. The primitive translational vectors $A_1$ ($a$, 0, 0) and $A_2$ (0, $b$, 0) are given in terms of two optimized lattice constants $a$ and $b$. The rectangular primitive cell contains four atoms with two nearly equaled Ga-P/In-P bonds ($b_1$, $b_2$) and three different angles ($\alpha_1$, $\alpha_2$, $\alpha_3$). Their positions are given by

$$(0,0,0), \quad (\frac{a}{2}, \frac{b}{2} - v\Delta z_2, \Delta z_1), \quad (\frac{a}{2}, \frac{b}{2}, \Delta z_1 + \Delta z_2), \quad (0, b - v\Delta z_2, \Delta z_2), \quad (1)$$

where $v = \sqrt{\left(\frac{b_2}{\Delta z_2}\right)^2 - 1}$, and $\Delta z_1$ and $\Delta z_2$ are buckling parameters which determine the total buckling $\Delta z_1 + \Delta z_2$.

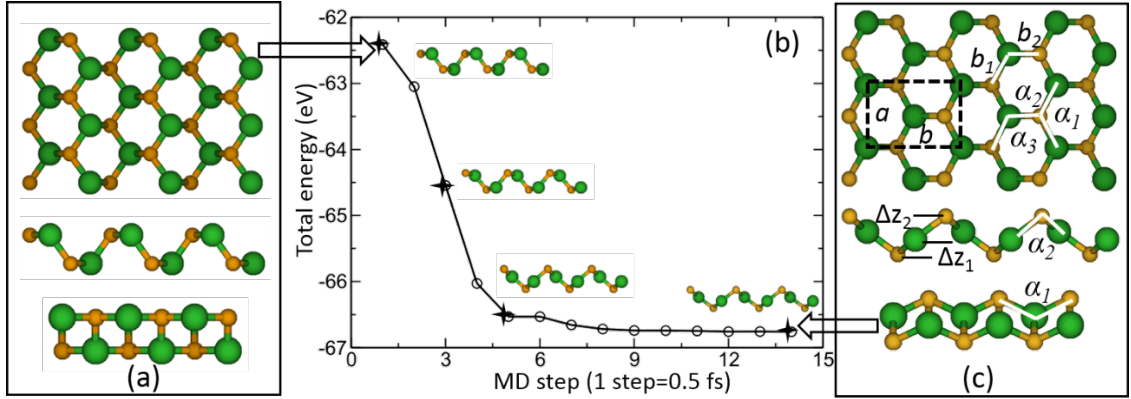

**Figure 2.** (color online) (a) Schematic illustrations of the top and two side views (seen from the front (middle) and from the left (bottom), respectively) of the proposed initial configurations of 2D GaP/InP monolayers (illustrated in Fig. 1 (c)). (b) The total energy of the 2D InP monolayer as a function of molecular dynamics (MD) steps. Inserts are the side views corresponding to some intermediate MD steps (marked with stars), demonstrating the evolution of the structure during the relaxation. The two open arrows indicate the structures at the initial (shown in (a)) and the final (shown in (c)) MD steps, respectively. (c) Schematic illustrations of the top  and two side views

(seen from the front (middle) and the left (bottom), respectively) of the stabilized high puckered orthorhombic 2D GaP/InP monolayers. The black-dashed box represents the rectangular primitive unit cell with lattice constants $a$ and $b$; $b_1$ and $b_2$ denote two types of Ga-P/In-P bonds; $\alpha_1$, $\alpha_2$, and $\alpha_3$ denote three types of angles; and $\Delta z_1$ and $\Delta z_2$ denote two buckling parameters, respectively. The yellow balls represent P atoms; and the green balls represent Ga/In atoms, respectively.

Optimized structural properties of the newly obtained 2D GaP/InP monolayer structures (referred as high puckered orthorhombic monolayer here after), their cohesive energies, and bandgaps (calculated with DFT [41, 42] and HSE06 [51] respectively) are listed in Tables 1-1 and 1-2, respectively. The corresponding values for the low buckled honeycomb 2D GaP/InP monolayers, as well as their bulk counterparts are also listed for comparisons. It is found that the two bond lengths ($b_1$ and $b_2$) are nearly equal (*i.e.*, 2.32 Å and 2.33 Å for GaP, and 2.53 Å and 2.54 Å for InP, respectively), and ~ 2.6 % shorter than their bulk counterparts, but ~ 1.3 % longer than their low buckled honeycomb counterparts. The three angles, on the other hand, are quite different and clearly demonstrate the anisotropic behaviors in these new high puckered orthorhombic structures. The first angle ($\alpha_1$) along the zigzag direction is close to the tetrahedral angle of $109^0$ (*i.e.*, $110.2^0$ for GaP and $108.6^0$ for InP). The second one ($\alpha_2$), charactering the dihedral angle, is $99.3^0$ for GaP and $97.5^0$ for InP, indicating a high buckling (*i.e.*, $\Delta z_1 + \Delta z_2$ is 2.14/2.46 Å for GaP/InP monolayers, which is about 5.94/4.56 times higher than those of the low buckled honeycomb GaP/InP monolayers). The third one ($\alpha_3$) on the semi-chair type of hexagon plane formed by the two Ga/In atoms and three P atoms is $124.6^0$ for GaP and $125.5^0$ for InP, slightly larger than the angle of flat hexagon ($120^0$). The characteristic of the three angles represent a mixture of the $sp^3$-like and $sp^2$-like hybridizations. Among the four $sp^3$-like orbitals, three of them bond with three nearest-neighbor atoms and the remaining one is perpendicular to the atomic layers of the 2D high puckered orthorhombic GaP/InP binary compounds. The π-like orbitals in $sp^2$-like hybridization, on the other hand, are perpendicular to the semi-chair type of hexagon planes.

In addition to their unique structural properties, another interesting point is their energetics. The cohesive energies per GaP/InP pair (defined as $E_c = E_{total} - E_{Ga/In} - E_P$, where $E_{total}$ is the total energy of the GaP/InP sheet per GaP/InP pair, $E_{Ga/In}$ and $E_P$ are the energies of single Ga/In and P atoms, respectively) are ~0.144 and 0.173 eV/pair lower than those of the low buckled honeycomb 2D GaP and InP structures (see the 6[th] row in Table 1-1 and 1-2), demonstrating that these newly predicted 2D GaP and InP monolayer structures are energetically preferential. Namely, the 2D GaP/InP monolayers prefer to maintain with high puckering, instead of low buckling. This tendency may be interpreted in terms of Jahn-Teller effect as the degeneracy at top valence band of the low buckled honeycomb 2D GaP/InP sheets [19] is removed in the new discovered high puckered orthorhombic 2D GaP/InP sheets by lowing the geometric symmetry, and the total energy is lowered.

**Table 1-1**. Optimized structural properties, cohesive energies, and DFT/HSE06 bandgaps of GaP allotropes, where $b_1$, $b_2$, $\alpha_1$, $\alpha_2$, and $\alpha_3$ are the bond lengths and angles of the high puckered orthorhombic GaP monolayer, respectively, as shown in Fig 2 (c). Numbers in parentheses (fourth column) are experimental results.

| GaP allotropes | High puckered orthorhombic monolayer | Low buckled honeycomb monolayer | Zinc Blende bulk |
|---|---|---|---|
| Lattice constants (Å) | 3.84 ($a$), 5.91 ($b$) | 3.92 | 5.53 (5.45)[a] |
| Bond lengths (Å) | 2.32 ($b_1$), 2.33 ($b_2$) | 2.29 | 2.39 (2.36)[a] |
| Bond angles (degree) | $110.2^0$ ($\alpha_1$), $99.3^0$ ($\alpha_2$), $124.6^0$ ($\alpha_3$) | $117.6^0$ | $109.0^0$ |
| Buckling parameters (Å) | 0.86 ($\Delta z_1$), 1.28 ($\Delta z_2$) | 0.36 | - |
| Cohesive energies (eV/pair) | -7.79 | -7.65 | -8.54 (-7.54)[b] |
| Band gaps (eV) | 1.97/2.89 | 1.54/2.51 | 1.52/2.37 (2.26)[a] |

**Table 1-2.** Optimized structural properties, cohesive energies, and DFT/HSE06 bandgaps of InP allotropes, where $b_1$, $b_2$, $\alpha_1$, $\alpha_2$, and $\alpha_3$ are the bond lengths and angles of the high puckered orthorhombic InP monolayer, respectively, as shown in Fig 2 (c). Numbers in parentheses (fourth column) are experimental results.

| InP allotropes | High puckered orthorhombic monolayer | Low buckled honeycomb monolayer | Zinc Blende bulk |
|---|---|---|---|
| Lattice constants (Å) | 4.13 ($a$), 6.33 ($b$) | 4.249 | 6.02 (5.88)[a] |
| Bond lengths (Å) | 2.53 ($b_1$), 2.54 ($b_2$) | 2.51 | 2.61 (2.55)[a] |
| Bond angles (degree) | $108.6^0$ ($\alpha_1$), $97.5^0$ ($\alpha_2$), $125.5^0$ ($\alpha_3$) | $115.6^0$ | $109.0^0$ |
| Buckling parameters (Å) | 0.98 ($\Delta z_1$), 1.48 ($\Delta z_2$) | 0.54 | - |
| Cohesive energies (eV/pair) | -7.06 | -6.89 | -7.78 (-6.88)[b] |
| Band gaps (eV) | 1.72/2.59 | 1.26/1.86 | 0.38/1.10 (1.34)[a] |

a: Ref. 49

b: Ref. 50

### 3.2 Phono dispersion

The dynamic stability of the high puckered orthorhombic 2D GaP and InP monolayers was examined from the analysis of the lattice vibrational modes using the combination of the phonon Boltzmann transport equation and the first-principles phonon calculations, as implemented in PHONOPY [52] code, which can directly use the force constants calculated by density functional perturbation theory. Here, a 4x4x1 (64 atoms) large supercell was used and the Brillouin zone is chosen as a Monkhorst-Pack grid of 12×16×1. Analogous to phosphorene [53], there are three acoustical and nine optical modes in these high buckled orthorhombic 2D GaP/InP monolayers. The atomic motions of lattice vibrational modes are illustrated in Fig. 3. Among them, six symmetric modes, i.e., $A_g^1$, $A_g^2$, $B_{1g}$, $B_{2g}$, $B_{3g}^1$, and $B_{3g}^2$ are Raman active modes based on the momentum conservation and the group theory [54].

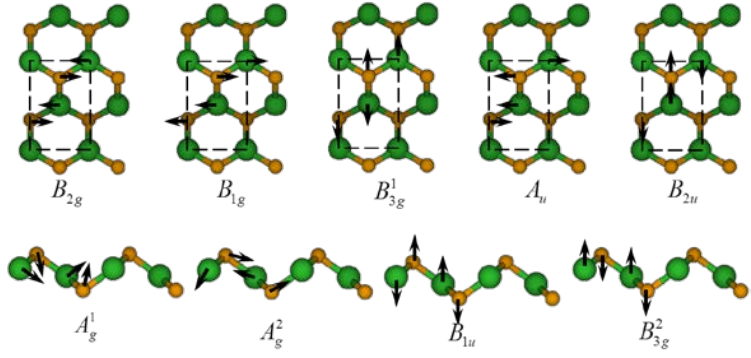

**Figure 3.** (color online) Schematic illustrations of the atomic motions of lattice vibrational modes of 2D high puckered orthorhombic GaP/InP monolayers. The upper panel illustrates $B_{2g}$, $B_{1g}$, $B_{3g}^1$, $A_u$, $B_{2u}$ modes viewed from the top, and the bottom panel illustrates $A_{1g}$, $A_g^2$, $B_{1u}$, $B_{3g}^2$ modes viewed from the side. The black-dashed boxes in the upper panel represent the primitive unit cell. The yellow balls represent P atoms; and the green balls represent Ga/In atoms, respectively.

Corresponding phonon dispersion spectra are presented in Figs. 4 (a) and (b), respectively. Both spectra have similar profile. There is no imaginary frequencies found in Brillouin zone, conforming that these high puckered orthorhombic 2D GaP/InP monolayers are energetically located at local minima on the Born-Oppenheimer surface and dynamically much stable as free standing sheets. Different from the phonon dispersion of acoustical modes in the low buckled honeycomb GaP/InP monolayers [19], the three acoustic modes in the high puckered orthorhombic GaP/InP monolayers are linear as the $k$-point closing to the Γ point. The infrared-active $B_{1u}$ modes (one of the out-of-plane optical modes) overlaps with the third acoustic mode (longitudinal acoustic mode) in the BZ along Y-S and S-X, and merges to $B_{1g}$ (one of the transverse optical modes) at Γ point. On the other hand, $B_{1g}$ and $B_{3g}^1$ (one of the longitudinal optical modes) modes merge each other between Y-S-X and begin to separate from Y and X points towards to the high symmetry Γ point. The remaining high frequency optical modes are doubly degenerate almost in the whole BZ, such as $A_u$ and $B_{2g}$, $A_g^2$ and $B_{2u}$, while $A_g^1$ and $B_{3g}^1$ are partially degenerated. Such degeneracies are lifted at Γ point. The frequencies of optical modes at Γ point are slightly lower than those in the low buckled honeycomb structures (see Fig. 3 in Ref. 19), which is mainly due to the longer bond lengths in the high orthorhombic structures. Like the profile of the phono dispersion in phosphorene [53], the speeds of sound along the Γ-Y direction are higher than those along the Γ-X direction, reflecting anisotropy in their elastic constants. Because of the heavier masses of Ga/In elements, the frequencies of the new allotropes of 2D GaP/InP monolayers are slightly lower than those of phosphorene monolayer [53] (*e.g.*, ~240 cm⁻¹ versus 365 cm⁻¹ in $A_g^1$ mode, and ~360 cm⁻¹ versus 420 cm⁻¹ near the $A_g^2$ and $B_{2g}$ modes, respectively).

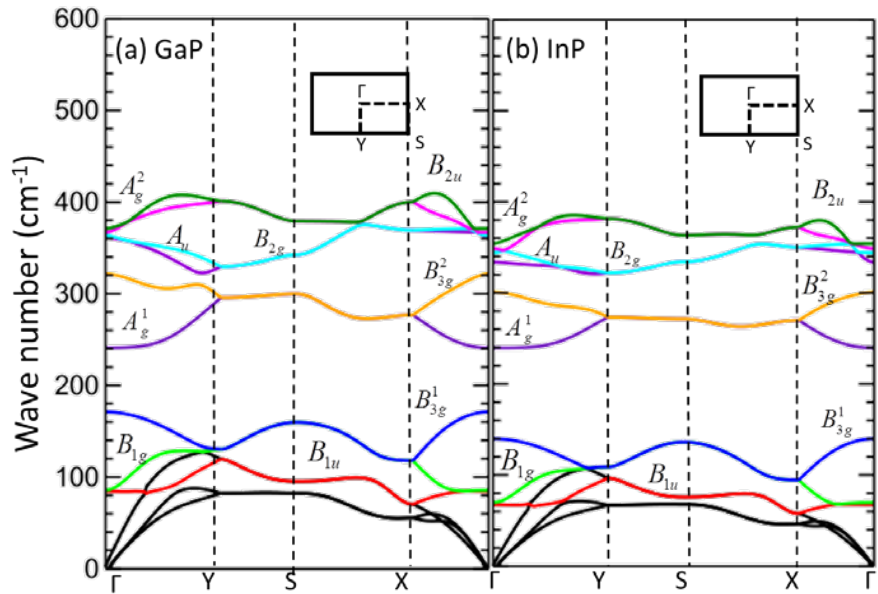

**Figure 4**. (color online) Calculated phonon dispersion spectra of the high puckered orthorhombic 2D GaP (a) and InP (b) monolayers. Their Brillouin zones are inserted. The nine optical modes are represented by color curves associated with symbols, and the three acoustical modes are represented by the black curves.

**3.3 Electronic band structures**

The anisotropic electronic properties of the high puckered orthorhombic 2D GaP/InP monolayers were systematically studied from electronic band structures and density of states (DOS) calculations, as presented in Figs. 5 (a) and (b). Apparently, their band structures demonstrate semiconductor behaviors. The high puckered orthorhombic 2D GaP monolayer has an indirect bandgap of 1.97 eV (or 2.89 eV in HSE06) between $\Gamma$ and Y points (Fig. 5 (a)). On the other hand, the high puckered orthorhombic 2D InP monolayer shows a direct band gap of 1.72 eV (or 2.59 eV in HSE06) at $\Gamma$ point (Fig. 5 (b)), maintaining the direct band gap nature of its bulk counterpart. The dispersions of charge carriers near the $\Gamma$ point also show anisotropic behavior, with higher speed towards to Y point and less speed towards to X point. Furthermore, it is found that calculated fundamental bandgaps are ~ 0.43/0.46 eV (or ~0.38/0.73 eV in HSE06) wider than those of the low buckled honeycomb GaP/InP monolayers and even ~ 0.45/1.34 eV (or ~ 0.52/1.49 eV in HSE06) wider than those of their bulk counterparts (see the 7[th] rows in Tables 1-1 and 1-2). These results indicate that the bandgaps in GaP/InP systems could be widened by reducing the dimension of the crystalline structures from 3D zinc blende structures to 2D high puckered orthorhombic structures.

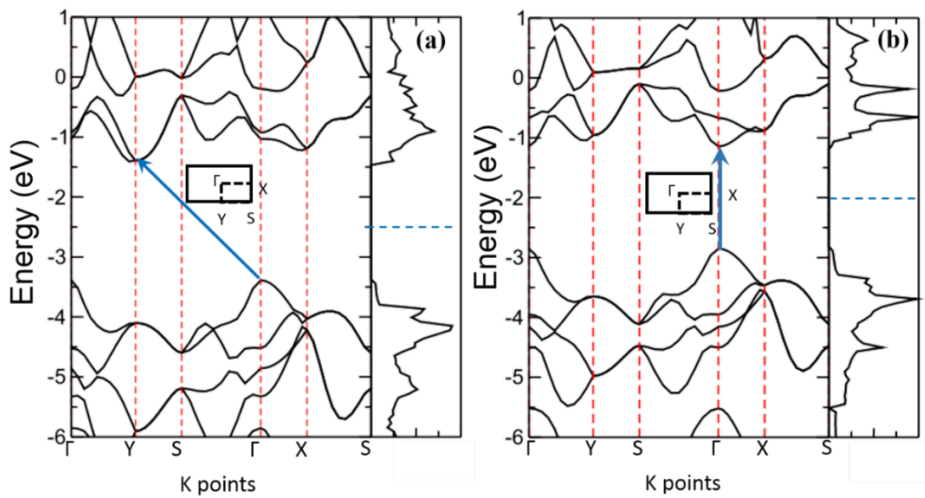

**Figure 5**. (color online) DFT band structures and DOSs of the high puckered orthorhombic GaP (a) and InP (b) monolayers. The Brillouin zones are inserted. The blue arrows in the band structures indicate directions from the tops of the valance bands to the bottoms of the conduction bands. The blue dash lines in the DOS (right panels in (a) and (b)) denote the fermi levels.

### 3.3 Tuning bandgap under strains

Most interesting finding is that the fundamental bandgaps of the high puckered orthorhombic 2D GaP and InP monolayers can be tuned by introducing different types of strain. Figs. 6 (a) and (b) presented their fundamental bandgaps ($E_g$) as a function of the in-plane strain σ (in percentage). The green open squares represent the bandgaps under the axial strain along the armchair direction; the red open diamonds represent the bandgaps under the axial strain along the zigzag direction; and the black open triangular represent the bandgaps under the biaxial strain, respectively. The corrected bandgaps calculated by HSE06 hybrid functions are also presented by corresponding solid symbols in Fig. 6. In the case of GaP sheet, the bandgap increases linearly with the increase of the strain along the armchair direction. However, it almost keeps unchanged in the case of InP sheet. No direct-indirect bandgap transition was found under the strain along the armchair direction (see the band structures under different strains in Figs. S1 (a) and (d) in Supplemental Material (SM)). While, a transition between indirect and direct bandgaps was found when the stress was added either along the zigzag direction (indicated by red dashed lines in Figs. 6 (a) and (b), and the band structures under strains in Figs. S1 (b) and (e) of the SM) or the biaxial (indicated by black dashed lines in Figs. 6 (a) and (b), and the band structures under strains in Figs. S1 (c) and (f) of the SM). Namely, the indirect bandgap nature (from the Γ to the Y points) in the high puckered orthorhombic 2D GaP monolayer can be tuned to the direct bandgap (at the Γ point) when the in-plan strain is over 2.5% (either under the strain along the zigzag direction by 2.6% or under the biaxial strain by 3%). The associated DFT bandgap increases from the 1.97 eV (with the zero strain) to 2.11 eV (along the zigzag) or 2.28 eV (by the biaxial) at the transition points and then gradually decreases after the transition points. On the other hand, the direct bandgap nature in the high puckered orthorhombic 2D InP monolayer at zero strain will stay under the strain along the armchair direction.

A transition from the direct to the indirect bandgap can only occur when the strain is negative, *i.e.*, -4% along the zigzag direction (see the red dotted-dash lines in Fig. 6 (b)) or -2.0 % under the biaxial strain (see the black dash lines in Fig. 6 (b)). The bandgap decreases as the decrease of the negative strain but increases as the increase of the positive strain, reaching a maximum at strain of ~ 4%. Even the bandgap slightly decreases after 4% of strain but it still keeps as large as 1.72 eV (~ 2.0 eV in HSE06) at large strains of 8%. These results clearly demonstrate that the high puckered orthorhombic GaP monolayer can be easily tuned to direct bandgap under a small elastic expansion (by ~3%). The high puckered orthorhombic InP monolayer, on the other hand, can maintain its direct bandgap nature even in a large non-elastic expansion (by ~8%), providing a very important fundamental guidance for functionally designing desired 2D nanoelectronic, optoelectronic, and photovoltaic devices through strain-induced bandgap engineering.

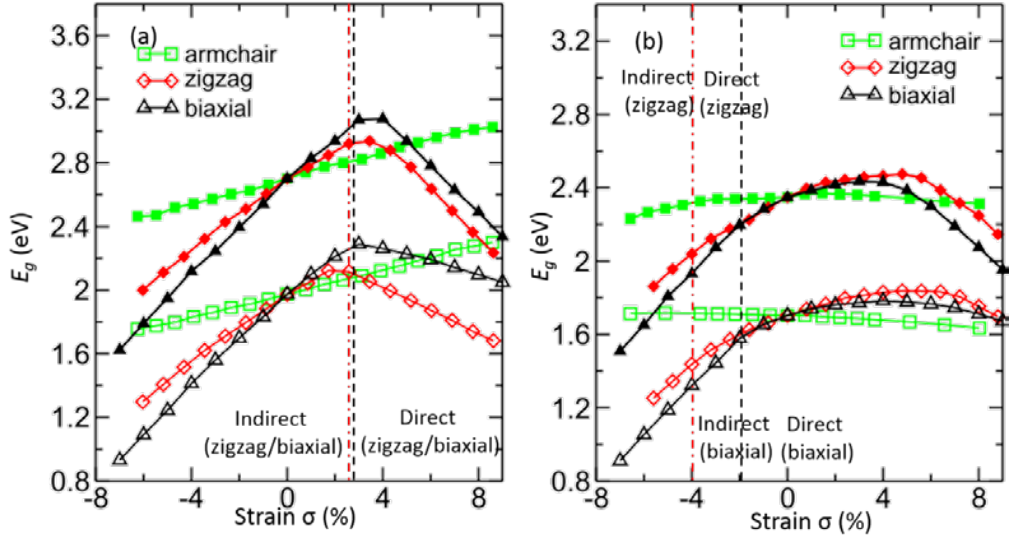

**Figure 6**. (color online) Calculated bandgaps ($E_g$) of the high puckered orthorhombic 2D GaP (a) and InP (b) monolayers as a function of the in-plane strains σ (in percentage), where, green open/solid squares denote the DFT/HSE06 bandgaps under the strain along the armchair direction, red open/solid diamonds denote the DFT/HSE06 bandgaps under the strains along the zigzag direction, and black open/solid triangles denote the DFT/HSE06 gaps under the biaxial strains, respectively. The red dotted-dash lines represent the direct/indirect transitions under the strain along the zigzag direction and black dash lines represent such transitions under the biaxial strains.

### 3.4 Anisotropic mechanical properties

The unique anisotropic structures of the high puckered orthorhombic GaP/InP monolayer binary compounds lead to the anisotropic behaviors in their mechanical properties. Figs. 7 (a)-(d)) show the strain energy density function Ψ as a function of the strain $\varepsilon$ along different directions. The strain energy density function Ψ is defined by

$$\Psi = \frac{E_{total}\left(\varepsilon\right) - E_{total}\left(0\right)}{A_{cell}}, \tag{1}$$

where $E_{total}\left(\varepsilon\right)$ and $E_{total}\left(0\right)$ are the total energy per atom with/without the strain, and $A_{cell}$ is the area of the unit cell at zero strain, respectively. It was found that the high puckered orthorhombic GaP and InP monolayers underwent an elastic expansion, a structural deformation, and then a structural broken processes as the strain increases (top and side views of the structures in the inserts of Figs. 7 (a)-(d)). Such processes strongly depend on the direction of the strain. The deformation was found when the strain is over 0.2 for the armchair expansion. However, it will easily occur if the expansion along the zigzag is over 0.12. For large strain ($\varepsilon > 0.3$), the armchair expansion will lead to bond broken along the armchair direction and the structures are destroyed to formed zigzag chains at large strain (top and side views of structures at the right side of the black dashed lines in Figs. 7 (a) and (c)). The zigzag expansion, on the other hand, will lead to a lattice change from a hexagonal ring to a rectangular ring when the strain is larger than 0.18 (top and side views of structures at the right side of the black dashed lines in Figs. 7 (b) and (d)).

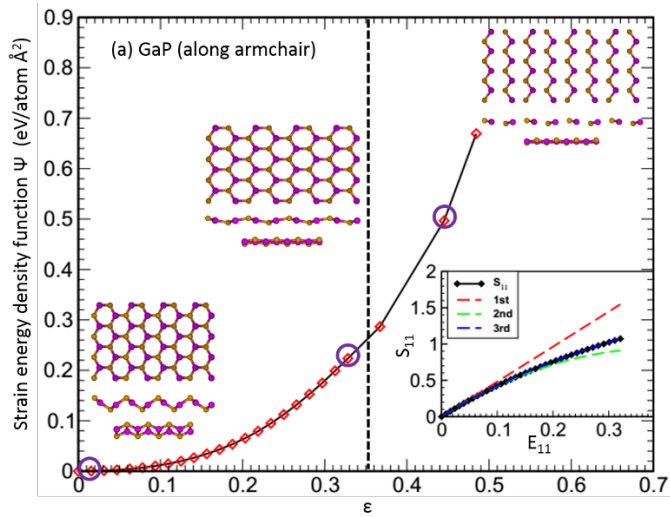

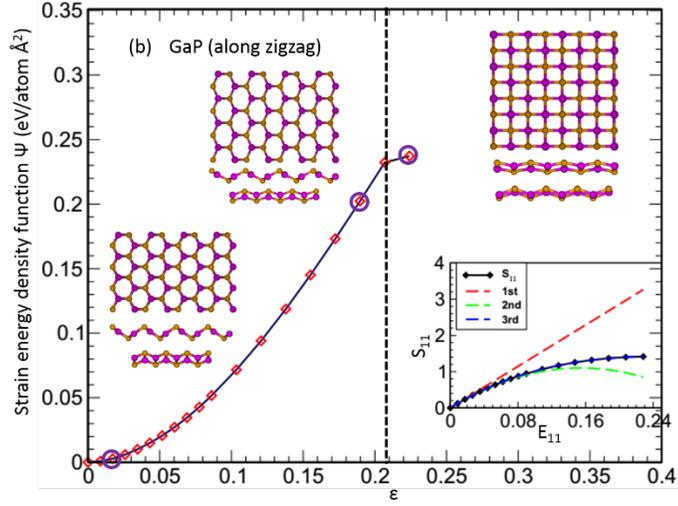

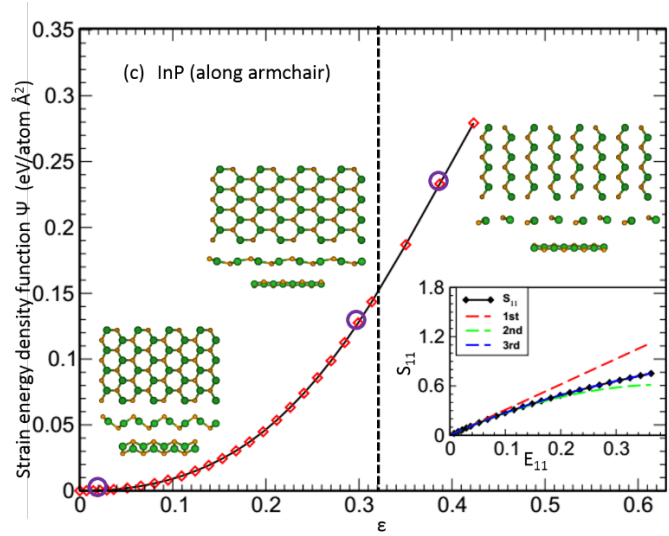

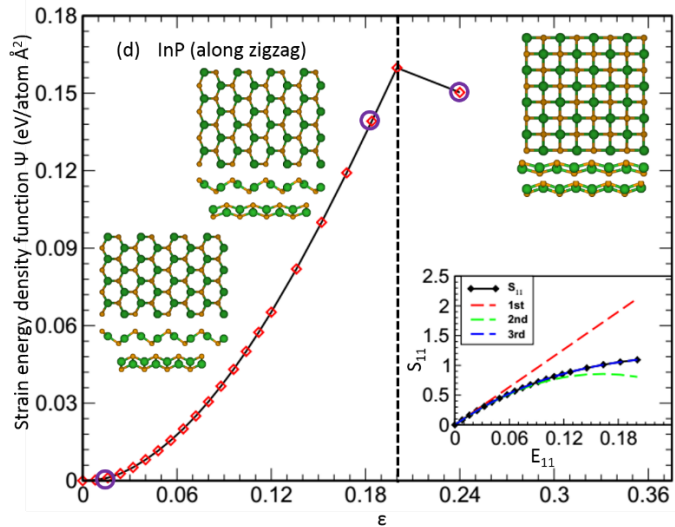

**Figure 7**. The strain energy density function $\Psi$ as a function of the strain $\varepsilon$ for the high puckered orthorhombic GaP along the armchair (a) and the zigzag (b) directions, respectively. Similarly, the strain energy density function $\Psi$ as a function of the strain $\varepsilon$ for the high puckered orthorhombic InP along the armchair and the zigzag directions are presented in (c) and (d), respectively. The red diamonds are the calculated values and the black curves are used to guide the results. The top and side views of the GaP/InP monolayers under different strains (marked by the blue circles on the curves) are shown in the inserts. The black dashed lines indicate the criteria of the strains after that the structures either broken or change to other type of lattice. The directional dependent 2D membrane stress along armchair/zigzag direction $S_{11}$ as a function of the uniaxial stretch in the given direction $E_{11,}$ are also presented in the lower right parts of the panels, where the red dashed lines represent the linear terms in $S_{11}$, the green dashed curves, the second order terms, and the blue dashed curves, the third order terms, respectively. The yellow balls represent P atoms; and the pink/green balls represent Ga/In atoms, respectively.

The directional dependence of the 2D membrane (tensile) stress along armchair/zigzag direction $S_{11}$ (analogous to the second Piola-Kirchhoff Stress in 3D [55]) was calculated from the derivative of the strain energy density function $\Psi$ with respect to the strain $\varepsilon$ (*i.e.*, $S_{11} = \dfrac{1}{1+\varepsilon}\dfrac{\partial \Psi}{\partial \varepsilon}$). It can be further expressed by Taylor series in terms of a uniaxial stretch in the given direction $E_{11}$ (here, $E_{11}(=\varepsilon + \dfrac{\varepsilon^2}{2})$ is the Green-Lagrange strain tensor elements [55] describing the physical and geometrical nonlinearity feature or large deformation). Namely, $S_{11} \approx Y^{2D} E_{11} + D^{2D} E_{11}^2 + F^{2D} E_{11}^3$, where $Y^{2D}$, $D^{2D}$, and $F^{2D}$ are the liner elastic constants (*i.e.*, the Young's modulus) and the high order effective nonlinear elastic moduli for 2D systems, respectively. The corresponding results are plotted in the inserts of Figs. 7 (a)-(d)). It was found that the elastic behavior (*i.e.*, the first order terms as indicated by the red-dashed lines) holds in the very small range of the strain (*i.e.*, < 0.05 for the armchair and < 0.025 for the zigzag directions, respectively). The second order nonlinear elastic behavior (indicated by the green-dashed lines) dominates in a large range of the strain (*i.e.*, $0.05 < \varepsilon < 0.18$ for the armchair direction and $0.025 < \varepsilon < 0.1$ for the zigzag direction). Estimated Young's moduli and high order effective nonlinear elastic moduli are listed in Table 2. They all show strong directional dependent nature. Especially, those constants along the zigzag direction are about one order in magnitude stronger than those along the armchair direction, indicating the strong anisotropic mechanical behavior in the high puckered orthorhombic GaP and InP monolayers. Compared to graphene (*e.g.*, the experimental value of graphene [56] is 340 Nm$^{-1}$ for $Y^{2D}$), they are softer, even along the zigzag direction. The poison ratios, on the other hand, were found to be close to zero in the linear elastic range (*i.e.*, $\varepsilon < 0.025$), almost independent of the direction of the strain exerted, mainly because of the reduction of the high buckling under the strain.

**Table 2**. Directional dependence of Young's moduli (3$^{rd}$ column) and the effective nonlinear elastic moduli (4$^{th}$ and 5$^{th}$ columns) of the high puckered orthorhombic GaP and InP monolayers.

| Binary compounds | direction | $Y^{2D}$ (Nm$^{-1}$) | $D^{2D}$ (Nm$^{-1}$) | $F^{2D}$ (Nm$^{-1}$) |
|---|---|---|---|---|
| GaP | armchair | 71.81 | -98.97 | 78.69 |
| GaP | zigzag | 228.82 | -741.340 | 763.39 |
| InP | armchair | 49.67 | -62.11 | 46.40 |
| InP | zigzag | 169.21 | -522.541 | 566.84 |

The present work also paved the way to synthesize these high puckered 2D GaP/InP monolayers by bulk truncating from the zinc blende crystalline structures (as shown in Figs. 1 (c)). To provide more fundamental guidelines for experimental synthesis, we conducted a various computational simulations and found that by substituting P atom alternatively with Ga/In atoms on phosphorene, the puckered lattice structures can automatically transform to the high puckered orthorhombic lattice (see Fig. S2 in SM). Our modeling results suggested another feasible way (*e.g*., by plasma assisted substituting CVD methods [57-59]) to realize the high puckered orthorhombic 2D GaP/InP binary compounds. Furthermore, compared with the band structure of phosphorene (DFT bandgap of 0.82 eV [60]; the HSE06 band gap of 1.0 eV [8], and experiment value of 1.45 eV [8]), it is also suggested that by substituting P atoms with Ga/In atoms on phosphorene, one can also engineer the energy gap of phosphide compounds.

## 4. Conclusion

The new allotropes of 2D GaP and InP monolayer structures with high puckered orthorhombic symmetry were predicted from the first-principle studies. Their stabilities are rigorously examined through structural optimization and lattice vibration mode calculations. They are energetically more stable than the previously predicted 2D GaP/InP sheets with low buckled honeycomb structures. They possess strong anisotropic and nonlinear mechanical properties. They are both semiconductor materials with the HSE06 functional band gaps of 2.89 eV and 2.59 eV, which are either direct at the $\Gamma$ point in the case of the InP monolayer, or indirect between $\Gamma$ and Y points in the case of the GaP monolayer. Most importantly, due to their anisotropic natures, their band gaps can be tuned by introducing strains either along the zigzag and armchair directions or biaxial. Especially, a transition between the indirect and direct band gaps can occur within a small strain range (less than $\pm$ 4%) either along the zigzag direction or under biaxial strain, providing intrigued hints for bandgap engineering.

**Acknowledgment**

We acknowledge computing resource support from the Cardinal Research Cluster at the University of Louisville.

**Supplemental Material Available:** Band structures of the GaP/InP monolayers under various strains (Figs. S1 (a)-(f)) and the schematic illustration of the fundamental guidance for synthesizing GaP/InP monolayers from phosphorene (Fig. S2).